\documentclass[
aps,
reprint,
prl,
superscriptaddress,
amsmath,
amssymb,
floatfix,
longbibliography,
noeprint
]{revtex4-2}

\usepackage{bm}
\usepackage{graphicx}
\usepackage{latexsym}
\usepackage{array}
\usepackage{amsfonts}
\usepackage[free-standing-units]{siunitx}
\usepackage[dvipsnames]{xcolor}
\usepackage{comment}

\usepackage{mathtools}  
\usepackage{makecell}   
\usepackage{hhline}     
\usepackage{amsthm}
\usepackage{amscd}

\renewcommand\vec{\boldsymbol}

\newcommand{\orcid}[1]{\href{https://orcid.org/#1}{\includegraphics[width=8pt]{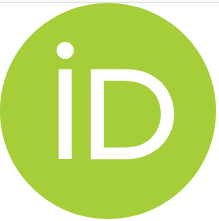}}}

\definecolor{orange}{rgb}{1,0.5,0}
\definecolor{goodGreen's}{rgb}{0.1,0.5,0}
\definecolor{goodred}{rgb}{0.7,0,0}
\usepackage[colorlinks,urlcolor=goodGreen's,citecolor=blue,linkcolor=goodred]{hyperref}
\begin{document}
\title{Josephson coupling through a magnetic racetrack}

\author{A. A. Mazanik \orcid{0000-0001-6389-8653}}
\email{andrei.mazanik@csic.es}
\affiliation{Centro de Física de Materiales (CFM-MPC) Centro Mixto CSIC-UPV/EHU, E-20018 Donostia-San Sebastián,  Spain}

\author{F. S. Bergeret \orcid{0000-0001-6007-4878}}
\affiliation{Centro de Física de Materiales (CFM-MPC) Centro Mixto CSIC-UPV/EHU, E-20018 Donostia-San Sebastián,  Spain}

\affiliation{Donostia International Physics Center (DIPC), 20018 Donostia-San Sebastián, Spain}

\begin{abstract}
We investigate the Josephson coupling between two superconducting electrodes connected by a ferromagnetic racetrack hosting a Bloch-like domain wall (DW). We show that the interplay between superconductivity and the DW leads to highly non-trivial spatial distributions of the supercurrent, including the formation of current loops and a strong sensitivity to the DW position and orientation. We further demonstrate that the Josephson critical  current $I_c$ can be efficiently controlled by the DW position along the racetrack, exhibiting pronounced variations and tunable $0$--$\pi$ transitions. These results provide clear design principles for superconducting racetrack devices and establish domain walls as a viable control element for readout schemes in racetrack memory architectures.
\end{abstract}
\maketitle

Magnetic racetrack memories are promising devices with exceptional speed and high-density storage capabilities \cite{parkin2008magnetic,ryu2013chiral,parkin2015memory}. Information is stored in the magnetic domain structure, specifically within the domains and the domain walls (DWs) that separate them. By applying an external current, these DWs can be displaced along the racetrack \cite{yamaguchi2004real}, leading to measurable changes in resistance of read-out junctions.

\begin{figure}[hbtp]
    \centering
\includegraphics[width=0.8\linewidth]{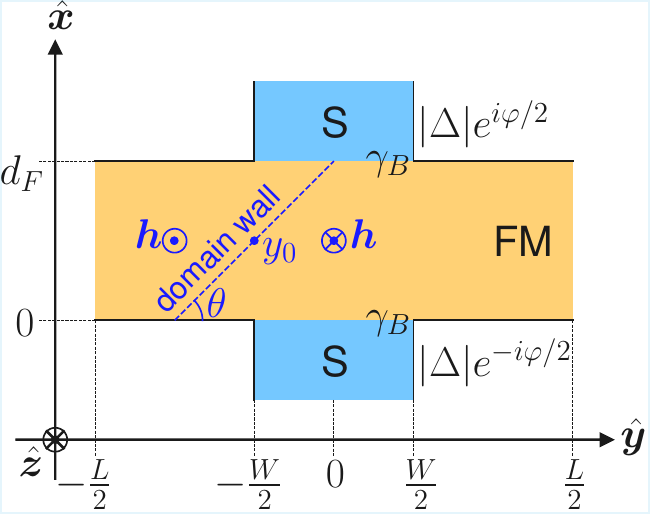}
    \caption{Josephson ferromagnetic racetrack under consideration. The domain wall is characterized by the spatial dependence of the exchange field, $\vec{h}(x,y)$ that is fixed by the wall  position $y_0$ and its tilt angle $\theta$. }
    \label{fig:setup}
\end{figure}

In parallel with the development of these devices, some works  have focused on combining racetrack architectures with superconducting elements \cite{rabinovich2018chirality,rabinovich2019resistive,hess2023josephson}. The key advantage -- and primary motivation -- of such hybrid systems lies in their potential for low energy consumption while maintaining high speed and scalability.
A typical setup studied is shown in Fig.~\ref{fig:setup}: two superconducting electrodes are coupled to the sides of a racetrack, thereby forming a local Josephson junction. The different memory states can then be read out through variations in the Josephson current. 
Beyond the control of domain wall motion itself, a central issue is how the wall position affects transport between the superconducting electrodes, in particular the Josephson current.


In this letter, we present a  study of the Josephson effect in a magnetic racetrack. Specifically, we analyze the device schematically shown in Fig.~\ref{fig:setup}, focusing on how the critical current of the junction depends on the position and properties of a domain wall, as well as other system parameters.  First, we compute the supercurrent distribution in the RT, which is highly non-trivial due to the spatially dependent Cooper-pair condensate (both singlet and triplet components). This leads to the emergence of supercurrent loops at the boundaries between strongly proximitized superconducting regions and bare ferromagnetic regions.
 Second, we show that the interaction between supercurrents and a DW is also highly non-trivial: when the DW is located inside the superconducting region, supercurrents tend to be attracted towards it, whereas when the DW is near the junction boundaries, supercurrents can be repelled from the DW. This behavior strongly depends on both the DW position and its orientation. Third, we find that DWs provide an efficient mechanism to control the critical current $I_c$, enabling tunable $0$--$\pi$ transitions as a function of the DW position, thus offering a practical route toward controllable Josephson devices.

\textit{System and Model.}
We consider the setup shown in Fig.~\ref{fig:setup}. It consists of a magnetic racetrack oriented along the $\hat{\vec{y}}$ axis and two superconducting electrodes of width $W$. The magnetization in the domains of the RT is assumed to be out of plane, with a Bloch-like domain wall separating two domains. The domain wall is characterized by the position of its center $y_0$, and its tilt, described by the angle $\theta$. The width of the RT is $d_F$. A similar setup was analyzed in Ref.~\cite{hess2023josephson} in the ideal ballistic limit. Here, we focus on a diffusive racetrack, which is more representative of metallic racetracks realized in experiments.

To describe transport in this system, we use the well-established Usadel equation. We assume that the proximity effect is weak, such that the equation can be linearized and takes the form of a diffusion-like equation for the superconducting condensate function $\check f$. In the RT, this equation has the form:
\begin{equation} \label{eq:Usadel}
    \nabla_k \check{J}_k = - \left[ \omega_n\check{\tau}_3 + i \vec{h}\cdot \vec{\sigma} \check{\tau}_3,\ \check{f}\right]\; .
\end{equation}
Here, $\check .$ describes matrices in spin-Nambu space, and $\sigma$ and $\tau$ are  the  Pauli matrices in these spaces.  $\omega_n$ is the Matsubara frequency, $\vec h$ is the spatially dependent exchange field vector that is assumed to be parallel to the local magnetization. The matrix current $\check J$ is defined as \cite{Kokkeler_2025}:
\begin{equation}
    \check{J}_k = - D \check{\tau}_3 \operatorname{sgn}{\omega_n} \nabla_k \check{f} + \frac{D \gamma  n_{ha}  \operatorname{sgn}{\omega_n}}{2} \left[ \sigma_a,\ \nabla_k \check{f} \right]\; ,
\end{equation}
where $D$ is the diffusion coefficient of the racetrack material, $\vec n_h = \vec{h}/h$, and $\gamma$ is the spin   polarization of the conducting electrons.

\begin{figure}[b]
    \centering
    \includegraphics[width=0.9\linewidth]{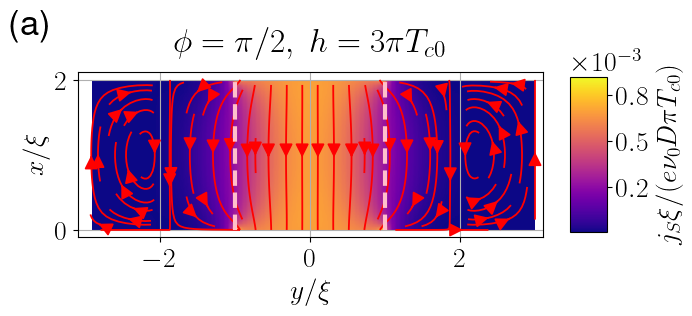}
    \includegraphics[width=0.9\linewidth]{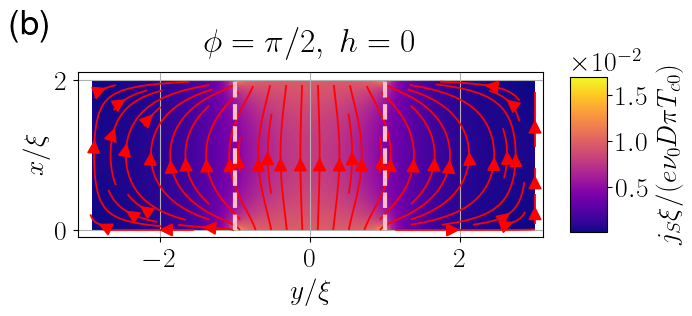}
    \caption{  Snapshots of supercurrent distributions for a racetrack without a domain wall for (a) $h=3\pi T_{c0}$ and (b) $h=0$ (normal-metal case). The region $ \vert y \vert < W/2$ is indicated by vertical pink dashed lines. Parameters: $\Gamma = 0$, $\gamma = 0.1$, $T = 0.1 T_{c0}$, $\gamma_B = 10\xi$, $W = 2\xi$, $L = 6\xi$, $d_F = 2\xi$.}
    \label{fig:supercurrent_no_DWs}
\end{figure}

To describe the superconducting proximity effect, we complement the above equations with boundary conditions at the superconducting contacts \cite{kuprianov1988influence}:
\begin{equation}\label{eq:BCs}
     \left. \check J_x\right|_{x=d_F,0}= \pm \frac{D}{2 R_b \sigma_F} \left[ \check{\tau}_3 \check{g}^{(u,l)}_{S} \check{\tau}_3  - \check{g}^{(u,l)}_{S} \right] {\cal F}(y)\; , 
\end{equation}
where ${\cal F}(y)=\Theta(y+W/2)-\Theta(y-W/2)$,  and the Green's functions in the lower and upper superconducting electrodes are given by 
\begin{equation}
    \check{g}^{(u,l)}_S = \frac{\omega_n}{\sqrt{\omega^2_n + \vert \Delta \vert^2}}\check{\tau}_3  + \frac{\vert \Delta \vert e^{\pm i \check{\tau}_3 \phi/2}}{\sqrt{\omega^2_n + \vert \Delta \vert^2}} \check{\tau}_2, 
\end{equation}
where $\vert \Delta \vert$ is the amplitude of the superconducting order parameter in the electrodes and $\phi$ is the phase difference between the superconductors.

For the DW, we assume the Bloch-like wall described by the following exchange vector $\vec{h}(x,y)$ entering Eq.~\eqref{eq:Usadel}: 
\begin{equation} \label{eq:h}
    \vec{h}(x,y) =h \begin{pmatrix}
         \sin \alpha(x,y) \sin\theta\\
          \sin \alpha(x,y) \cos\theta\\
          \cos \alpha(x,y)
    \end{pmatrix}
\end{equation}
where 
\begin{equation}\label{eq:alpha_xy}
    \alpha(x,y)=2 \arctan \exp \left[\sin\theta \frac{y - y_0}{w}  + \cos\theta \frac{x - d_F}{w}  \right]. 
\end{equation}
Here, $w$ corresponds to the width of the DW.  Eqs.~\eqref{eq:h} and \eqref{eq:alpha_xy} describe a Bloch-like domain wall.

In the following, we solve Eqs.~\eqref{eq:Usadel}--\eqref{eq:alpha_xy} numerically  via the finite element method using the FEniCSx library \cite{BarattaEtal2023,ScroggsEtal2022,BasixJoss2022,AlnaesEtal2014} and determine the supercurrent  density from the expression 
\begin{equation} \label{eq:supercurrent}
\begin{aligned}
    &\vec{j}_S = \frac{i}{2} e \pi \nu_0  DT \sum_{\omega_n} \operatorname{tr}\left\{ \check{\tau}_3  \left[ \frac{1}{2}\left( \check{f}\nabla \check{f}  -  \nabla \check{f}  \check{f} \right) - \right.\right.\\
    &\qquad - \frac{\gamma n_{ha}}{4} \left\{\check{\tau}_3 \sigma_a,\ \check{f} \nabla \check{f} - \nabla \check{f} \check{f} \right\} -\\  
    &\qquad \left.\left. - \frac{\gamma n_{ha}}{4}\left\{ \check{f}\sigma_a + \sigma_a \check{f},\ \check{\tau}_3 \nabla \check{f}\right\} \right] \right\},
\end{aligned}
\end{equation}
where $T$ is the temperature of the system, and $\nu_0$ is the density of states per spin at the Fermi 
level.

\textit{Results.}  We begin by considering the racetrack in the absence of a domain wall. As shown in Figs.~\ref{fig:supercurrent_no_DWs}(a) and (b), the supercurrent distribution already becomes non-trivial in the presence of an exchange field. In particular, for $h = 3\pi T_{c0}$, where $T_{c0}$ denotes the BCS critical temperature of the superconducting electrodes, the interfaces between the Josephson junction   ($|y| < W/2$) and the racetrack regions ($|y| > W/2$) act as sources of supercurrent inhomogeneity. This leads to the formation of supercurrent vortices near the edges of the junction, as shown in Fig.~\ref{fig:supercurrent_no_DWs}(a), which are absent in the $h = 0$ case [Fig.~\ref{fig:supercurrent_no_DWs}(b)].
Physically, we attribute the emergence of these current loops to the spatially nonuniform proximity effect near the edges of the junction region, which produces a nonuniform phase structure of the condensate and, consequently, circulating supercurrents.


\begin{figure}[t]
    \centering
    \includegraphics[width=0.9\linewidth]{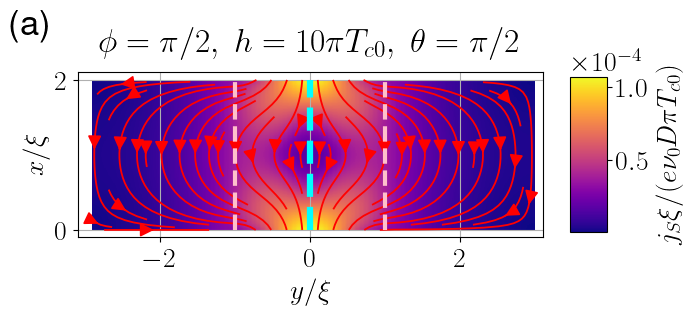}
    \includegraphics[width=0.9\linewidth]{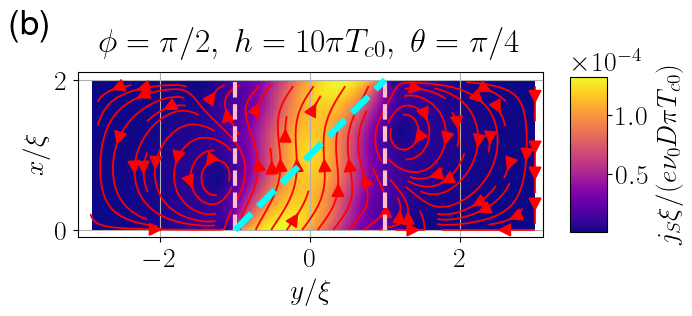}
    \includegraphics[width=0.9\linewidth]{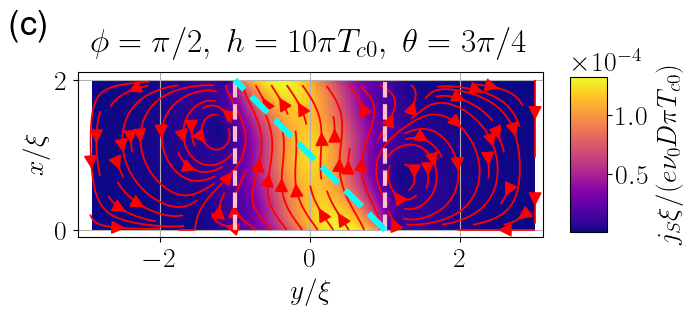}
    \caption{Snapshots of supercurrent distributions when the domain wall is in the middle of the junction, $y_0 = 0$.  The region $ \vert y \vert < W/2$ is indicated by vertical pink dashed lines. Cyan dashed lines show the positions of the DW. Parameters: $w = 0.5\xi$, $\Gamma = 0$, $\gamma = 0.1$, $T = 0.1 T_{c0}$, $\gamma_B = 10\xi$, $W = 2\xi$, $L = 6\xi$, $d_F = 2\xi$. }
    \label{fig:DW_middle}
\end{figure}

\begin{figure}[t]
    \centering
    \includegraphics[width=0.9\linewidth]{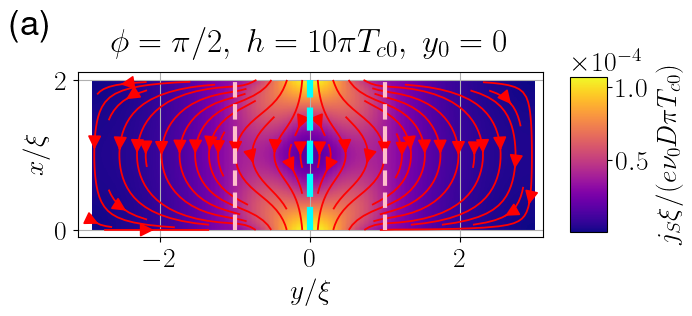}
    \includegraphics[width=0.9\linewidth]{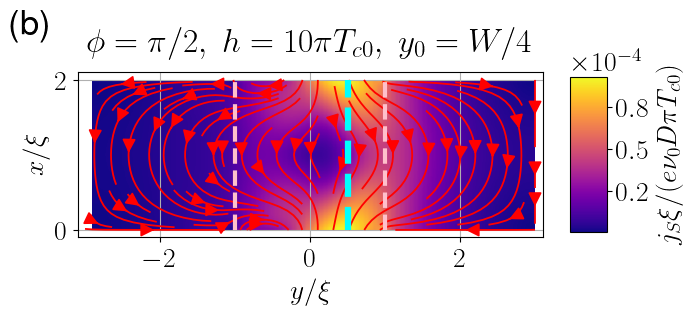}
    \includegraphics[width=0.9\linewidth]{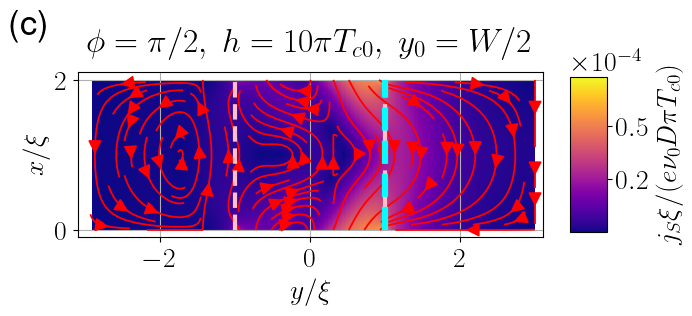}
    \includegraphics[width=0.9\linewidth]{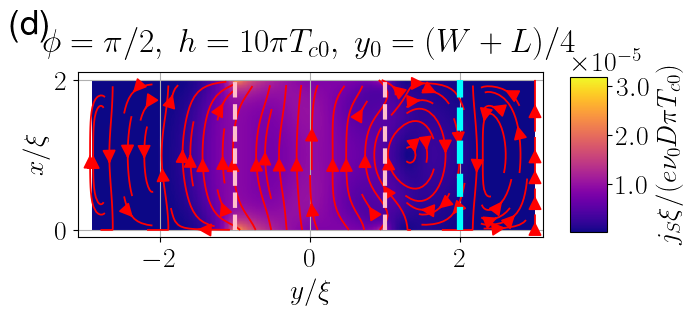}
    \caption{Snapshots of supercurrent distributions for the domain wall moving along the racetrack, $\theta= \pi/2$. The region $ \vert y \vert < W/2$ is indicated by vertical pink dashed lines. Cyan dashed lines show the positions of the DW. Parameters: $w = 0.5\xi$, $\Gamma = 0$, $\gamma = 0.1$, $T = 0.1 T_{c0}$, $\gamma_B = 10\xi$, $W = 2\xi$, $L = 6\xi$, $d_F = 2\xi$.  }\label{fig:y_0effect}
\end{figure}

When a DW is placed near or into the superconducting region of the JJ, then the supercurrent interacts with the DW in a way that strongly depends on the DW position. In particular, when the DW is in the middle of the junction, $y_0 = 0$, then the supercurrent prefers to flow along the DW despite the orientation of the DW with respect to the supercurrent electrodes; see Figs.~\ref{fig:DW_middle}(a)--(c). This behavior is understood in terms of the formation of short- and long-range triplet correlations due to the DW \cite{bergeret2001long,bergeret2001josephson,bergeret2005odd}, which is sensitive to the geometric details of the magnetic texture and junction \cite{Fominov2007DomainWallsJJ,Pugach2011,Robinson2012,bergeret2013singlet}.  We also see that when the DW is located at the center of the junction and is tilted with respect to the superconducting electrodes, it forms a supercurrent transport channel that is more efficient than that produced by a DW oriented perpendicular to the electrodes. The emergence of such efficient channels for tilted DWs of finite width extends the conclusions of \cite{Pugach2011}, where it was shown that sharp domain walls perpendicular to the electrodes do not support effective supercurrent transport.

When the DW position $y_0$ is changed along the racetrack, then the supercurrent may be as attracted, as repelled from the DW, see Figs.~\ref{fig:y_0effect}(a)--(d). We attribute the regime in which the supercurrent is repelled from the domain wall to the interplay between the triplet formation triggered by the DW and the edge current distribution shown in Fig.~\ref{fig:supercurrent_no_DWs}(a). We note that the mechanism of the supercurrent-texture interaction  is given only by ferromagnetic exchange in our model; thus it is different from the mechanism via stray fields \cite{Andriyakhina2022}.

\begin{figure}
    \centering
    \includegraphics[width=0.9\linewidth]{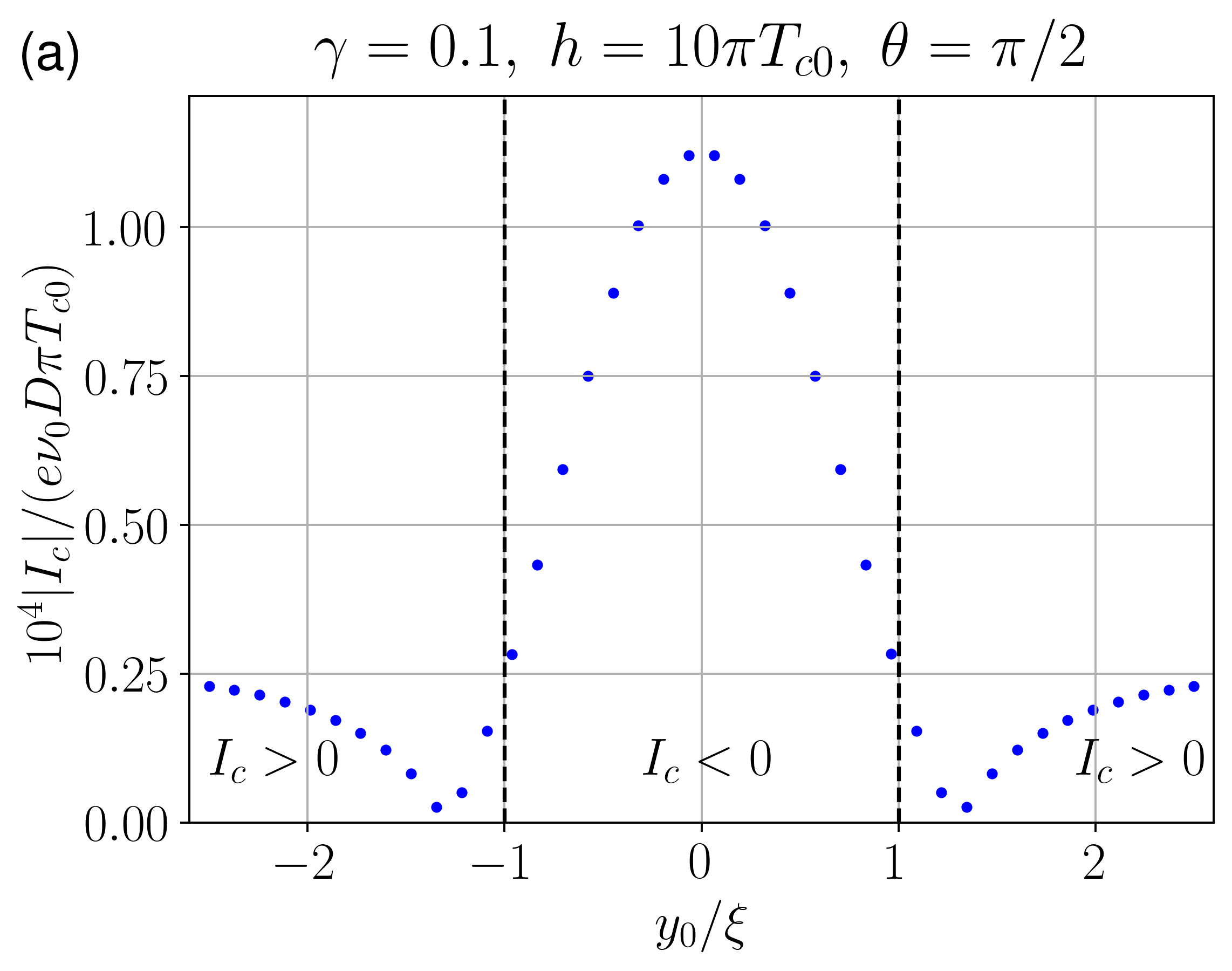}
    \includegraphics[width=0.9\linewidth]{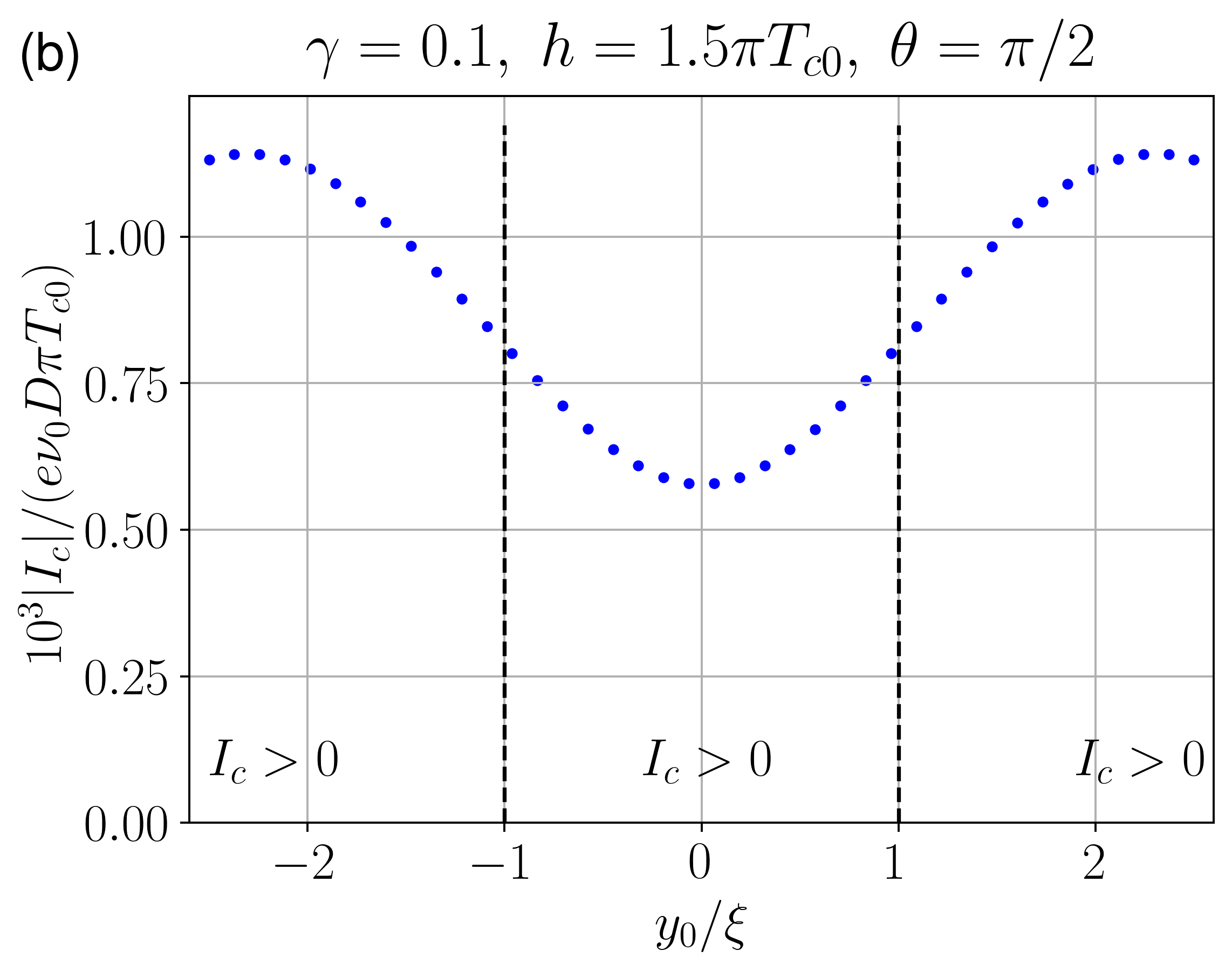}
    \caption{Critical current dependencies on the DW position along the racetrack, $I_c(y_0)$, for two different values of the exchange field. Vertical dashed lines represent the region $\vert y_0 \vert < W/2$. The simulation parameters are as follows: $w = 0.5\xi$, $\Gamma = 0$, $\gamma = 0.1$, $T = 0.1 T_{c0}$, $\gamma_B = 10\xi$, $W = 2\xi$, $L = 6\xi$, $d_F = 2\xi$}
    \label{fig:I_c_y0}
\end{figure}

Finally, we compute how the critical current of the JJ, i.e., the maximum current that can flow between the superconductors, depends on the DW position along the racetrack, $y_0$, for  a strong, $h = 10 \pi T_{c0}$, and a weak, $h = 1.5 \pi T_{c0}$, ferromagnet. The results of this calculation are shown in Figs.~\ref{fig:I_c_y0}(a) and (b). We see that the interaction between the supercurrent and the magnetic texture discussed above results in a pronounced variation of the critical current during the change of $y_0$. When the DW enters the superconducting region ($\vert y \vert < W/2$), the critical current is reduced for both values of the exchange field. For $h = 10 \pi T_{c0}$, this suppression becomes sufficiently strong to induce $0$--$\pi$ transitions driven by the DW motion. It can be understood as follows. When the polarization of the ferromagnet is small, $\gamma \ll 1$, the supercurrent  Eq.~\eqref{eq:supercurrent}  injected into the racetrack through the lower electrode can be expressed as $I_S = \int_{-W/2}^{W/2} dy\, j_{Sx}(y) \propto \int_{-W/2}^{W/2} dy \sum_{\omega_n} \operatorname{Im}\left[ f^\star_s \partial f_s - f^\star_{ti} \partial_x f_{ti} \right]$, where $f_s$ and $f_{ti}$ [$i = x,y,z$] denote the singlet and triplet components of $\check{f}$, respectively. Since the singlet component  $f_s$ is purely imaginary and the  triplet components $f_{ti}$ are real, their increase caused by the DW results in a decrease of the critical current. This indicates that the DW acts as a source of singlet-to-triplet conversion, which becomes increasingly pronounced with larger exchange fields and causes the $0$--$\pi$ transitions. One can envision a direct application of such a $0$--$\pi$ transition as the two states of the racetrack memory, controlled by the position of the DW.

\textit{Conclusions.} In this work, we have investigated the supercurrent flow through a Josephson junction containing a ferromagnetic racetrack hosting a Bloch-like domain wall.  We have found that the supercurrent interacts with the domain wall in the following way. When the wall is located inside the region in which the racetrack is connected to  the electrodes, then the supercurrent is attracted to the wall and mainly prefers to flow along it. However, when the domain wall approaches the edges of the superconducting region, this attraction is replaced by repulsion.  We have shown that this interplay between the magnetic texture, supercurrent, and geometry is accompanied by a pronounced tuning of the critical current as a function of the wall position along the racetrack. We believe that these results will boost the creation of superconducting racetrack memory and, in particular, its reading procedure.  

\textit{Acknowledgments.}
We thank Tim Kokkeler for useful discussions, and the  financial support from the Spanish MCIN/AEI/10.13039/501100011033 
through the grant PID2023-148225NB-C31, and  the
European Union’s Horizon Europe research and innovation program under grant agreement No. 101130224 (JOSEPHINE).

\bibliographystyle{apsrev4-2}
\bibliography{refs}
\end{document}